%% file: main.tex
\documentclass[aps,prc,preprint,showpacs,showkeys,superscriptaddress]{revtex4-1}  
\usepackage{amssymb}   
\usepackage{graphicx}  
\usepackage{bm}        
\usepackage{amsmath}
\usepackage{dcolumn}   
\usepackage{xspace}    
\usepackage{color}     
\usepackage{url}       
\usepackage{siunitx}   
\usepackage{subfig}
\usepackage{verbatim}
\usepackage{appendix}

\setkeys{Gin}{clip=true,keepaspectratio=true} 
\begin{document}
   \title{Systematic and Statistical Uncertainties of the Hilbert-Transform Based High-precision FID Frequency Extraction Method}

\input{author_list} 
   \begin{abstract}
   Pulsed nuclear magnetic resonance (NMR) is widely used in high-precision magnetic field measurements. The absolute value of the magnetic field is determined from the precession frequency of nuclear magnetic moments. The Hilbert transform is widely used to extract the phase function from the observed free induction decay (FID) signal and then its frequency. In this paper, a detailed implementation of a Hilbert-transform based FID frequency extraction method is described. How artifacts and noise level in the FID signal affect the extracted phase function are derived analytically. A method of mitigating the artifacts in the extracted phase function of an FID is discussed. Correlations between noises of the phase function samples are studied for different noise spectra. We discovered that the error covariance matrix for the extracted phase function is nearly singular and improper for constructing the $\chi^2$ used in the fitting routine. A down-sampling method for fixing the singular covariance matrix has been developed, so that the minimum $\chi^2$-fit yields properly the statistical uncertainty of the extracted frequency. Other practical methods of obtaining the statistical uncertainty are also discussed.
   

   
\end{abstract}

   \keywords{FID, high-precision magnetometer, frequency extraction, Hilbert transform, uncertainty analysis}

   \maketitle
   \input{Introduction.tex}

   \input{Method.tex}
   \input{Artifact.tex}

   \input{Noise.tex}
   \input{Conclusion.tex}
   \section{Acknowledgements}
   This work was supported by the U.S. Department of Energy (DOE), Office of Science under contracts DE-AC02-06CH11357 (Argonne National Laboratory), DE-FG02-88ER40415 (University of Massachussetts), and DE-FG02-97ER41020 (University of Washington), and National Science Foundation (NSF), Division of Physics Award No. 1807266 (University of Kentucky) and Award No. 1812314 (University of Michigan). This work was also supported in part by the National Natural Science Foundation of China Grant No. 11975153 (Shanghai Jiao Tong University).
   
   This document was prepared by the Muon $g-2$ collaboration using the resources of the Fermi National Accelerator Laboratory (Fermilab), a U.S. Department of Energy, Office of Science, HEP User Facility. Fermilab is managed by Fermi Research Alliance, LLC (FRA), acting under Contract No. DE-AC02-07CH11359.

   \input{Appendix.tex}

   \bibliographystyle{unsrt}
   \bibliography{References}{}  
\end{document}

%% file: author_list.tex
\author{Ran Hong}
\email{rhong@anl.gov}
\affiliation{Argonne National Laboratory, Lemont, IL, USA}
\affiliation{University of Kentucky, Lexington, KY, USA}

\author{Simon Corrodi}
\affiliation{Argonne National Laboratory, Lemont, IL, USA}

\author{Saskia Charity}
\affiliation{Fermi National Accelerator Laboratory, Batavia, IL, USA}

\author{Stefan Bae\ss{}ler}
\affiliation{University of Virginia, Charlottesville, VA, USA}
\affiliation{Oak Ridge National Lab, Oak Ridge, TN, USA}

\author{Jason Bono}
\affiliation{Fermi National Accelerator Laboratory, Batavia, IL, USA}

\author{Timothy Chupp}
\affiliation{University of Michigan, Ann Arbor, MI, USA}

\author{Martin Fertl}
\affiliation{University of Washington, Seattle, WA, USA}
\affiliation{Johannes Gutenberg-Universit\"at Mainz, Mainz, Germany}

\author{David Flay}
\affiliation{University of Massachusetts, Amherst, MA, USA}

\author{Alejandro Garc\'{i}a}
\affiliation{University of Washington, Seattle, WA, USA}

\author{Jimin George}
\affiliation{University of Massachusetts, Amherst, MA, USA}

\author{Kevin Louis Giovanetti}
\affiliation{James Madison University, Harrisonburg, VA, USA}

\author{Timothy Gorringe}
\affiliation{University of Kentucky, Lexington, KY, USA}

\author{Joseph Grange}
\affiliation{Argonne National Laboratory, Lemont, IL, USA}
\affiliation{University of Michigan, Ann Arbor, MI, USA}

\author{Kyun Woo Hong}
\affiliation{University of Virginia, Charlottesville, VA, USA}

\author{David Kawall}
\affiliation{University of Massachusetts, Amherst, MA, USA}

\author{Brendan Kiburg}
\affiliation{Fermi National Accelerator Laboratory, Batavia, IL, USA}

\author{Bingzhi Li}
\altaffiliation[Also at ]{Shanghai Key Laboratory for Particle Physics and Cosmology, Shanghai, China}
\altaffiliation[Also at ]{Key Lab for Particle Physics, Astrophysics and Cosmology (MOE), Shanghai, China}
\affiliation{Argonne National Laboratory, Lemont, IL, USA}
\affiliation{ Shanghai Jiao Tong University, Shanghai, China}

\author{Liang Li}
\altaffiliation[Also at ]{Shanghai Key Laboratory for Particle Physics and Cosmology, Shanghai, China}
\altaffiliation[Also at ]{Key Lab for Particle Physics, Astrophysics and Cosmology (MOE), Shanghai, China}
\affiliation{ Shanghai Jiao Tong University, Shanghai, China}

\author{Rachel Osofsky}
\affiliation{University of Washington, Seattle, WA, USA}

\author{Dinko Po\v{c}ani\'c}
\affiliation{University of Virginia, Charlottesville, VA, USA}

\author{Suvarna Ramachandran}
\affiliation{Argonne National Laboratory, Lemont, IL, USA}

\author{Matthias Smith}
\affiliation{University of Washington, Seattle, WA, USA}

\author{Herbert Erik Swanson}
\affiliation{University of Washington, Seattle, WA, USA}

\author{Alec Tewsley-Booth}
\affiliation{University of Michigan, Ann Arbor, MI, USA}

\author{Peter Winter}
\affiliation{Argonne National Laboratory, Lemont, IL, USA}

\author{Tianyu Yang}
\affiliation{University of Michigan, Ann Arbor, MI, USA}

\author{Kai Zheng}
\affiliation{Argonne National Laboratory, Lemont, IL, USA}

\collaboration{The Muon $g-2$ Collaboration}
\homepage{http://muon-g-2.fnal.gov/collaboration.html}

%% file: Introduction.tex
\section{Introduction}

Proton nuclear magnetic resonance (NMR) magnetometers are widely used in high precision magnetic field measurements \cite{hartmann:1972}. The magnetic field magnitude $B$ is determined by measuring the proton spin precession angular frequency $\omega_{s} = \gamma B$ using a proton-rich material, where $\gamma$ is the gyro-magnetic ratio of a proton. 
The magnetization of the detection material is aligned with the magnetic field $\bold{B}$ in thermal equilibrium. 
In the pulsed NMR measurement scheme, a pulsed oscillating magnetic field ($\pi/2$-pulse) transverse to $\bold{B}$ with an angular frequency near $\omega_{s}$ is generated by a coil surrounding the detection material, which tips the magnetization into the transverse plane. 
After the $\pi/2$-pulse, the precessing magnetization generates an oscillating signal that can be picked up in the same coil, amplified, and detected. 
The signal amplitude decays due to the relaxation of the magnetization.
Therefore, the detected signal of the pulsed NMR is referred to as the free induction decay (FID). FID signals can be analyzed by hardware spectrometers, or be digitized and stored so that more sophisticated analysis algorithms can be performed by a computer or an embedded system. Often the FID signal is mixed with a sinusoidal reference with an angular frequency $\omega_{R} \approx \omega_{s}$. The mixed signal is then passed through a low-pass filter that keeps the $|\omega_{s} - \omega_{R}|$ component. This reduces the sampling frequency requirement, data rate and readout noise. 

Pulsed proton NMR magnetometers typically have a precision better than  1~part-per-million (ppm), and they have already been used in many nuclear physics and high-energy physics experiments \cite{PRIGL1996118,FEI1997349}. For example, the Muon $g-2$ Experiment \cite{Grange:2015fou} at Fermilab uses pulsed NMR probes to measure the magnetic field in the storage ring, and the uncertainty budget for FID frequency extraction is 10~part-per-billion (ppb). To achieve such a high precision, it is critical to evaluate the systematic and statistical uncertainties introduced by the read-out system. Due to saturation effects of the amplifiers, imperfections of the mixer, and pedestal instabilities of the Analog to Digital Converter (ADC), the FID signal is distorted and a non-zero baseline is added to the signal. Understanding how biases are introduced through these effects quantitatively will help in determining specifications of components when designing an NMR magnetic field measurement system, and estimating the systematic uncertainties when they are irreducible. On the other hand, noises introduced by the electronics lead to statistical uncertainty in the FID frequency measurement, and it is important to understand this relationship in order to fully describe the uncertainty of the FID frequency measurement.


Many methods for improving the accuracy and resolution of FID frequency measurements have been developed recently for medical applications \cite{Lu:1997} and weak field measurements \cite{Dong:2016}. For high-energy physics experiments that require sub-ppm level uncertainties, one challenge is to make measurements in regions with a significant field inhomogeneity. 
In an inhomogeneous magnetic field, the nuclear spin precession frequencies vary across the volume of the detection material, and the superposition of signals with different frequencies results in an FID with a broadened and complicated frequency spectrum. Using advanced frequency extraction algorithms, the average NMR frequency sensed by the probe can be resolved with a precision better than its frequency-domain line width. The FID analysis method \cite{Cowan_1996} developed by Cowan {\em et. al.} relates the average NMR frequency to the derivative of the FID phase function, which can be extracted using several methods like zero-crossing counting and the Hilbert-transform method described in Sec.~\ref{sec:method}. 
Noise and error analyses have been performed on proton-NMR magnetometers using zero-crossing based frequency extraction methods \cite{Denisov_2014,Liu_2017}. However, the noise spectrum in the phase function and the statistical uncertainty of the FID frequency extracted using the Hilbert-transform method have not been thoroughly investigated. The goal of this study is to quantify the systematic uncertainties caused by artifacts, and develop a method for determining the statistical uncertainty of the FID frequency extraction. A detailed implementation of the phase function extraction using the Hilbert transform and Cowan's method for frequency determination are described in Sec.~\ref{sec:method}. The mechanism of how the discrete Hilbert transform and artifacts like the signal distortion and baseline affect the phase function of an FID is presented in Sec.~\ref{sec:signal_artifacts}. The systematic uncertainties caused by these effects and a mitigation method are discussed as well. Furthermore, the noise spectrum in the phase function and the statistical uncertainty for Cowan's method are discussed in Sec.~\ref{sec:noise}.

%% file: Method.tex
\section{FID Frequency Extraction Method}
\label{sec:method}

In an inhomogeneous magnetic field, the general form of an FID resulting from the superposition of signals with different frequencies can be modeled as 
\begin{align}
f(t)&=N\exp\left(-\frac{t}{T_{2}}\right)\int_{-\infty}^{+\infty}g(\omega)\exp(i(\omega t + \phi_0))d\omega,\label{eq:fid_nonuniform}
\end{align}
where $N$ is a normalization constant, $\phi_{0}$ is the initial phase, and $T_{2}$ is the intrinsic transverse relaxation time constant of the detection material \cite{Slichter:1990}. The spectrum density function $g(\omega)$ is normalized so that $\int_{-\infty}^{+\infty}g(\omega)d\omega=1$, and $g(\omega)d\omega$ is proportional to the amplitude of the signal with an angular frequency within the range $(\omega,\omega+d\omega)$.
The function $f(t)$ is complex, and the measured signal is its real part $f_r$. 
The FID function $f(t)$ can be expressed in the form of a general complex function
\begin{align}
    f(t)=A(t)\exp(i\Phi(t)),
    \label{eq:general_complex}
\end{align}
where $A(t)$ and $\Phi(t)$ are real. According to Ref.~\cite{Cowan_1996}, the average NMR frequency $\bar{\omega}$ weighted by $g(\omega)$ can be determined by calculating the derivative of $\Phi(t)$ at $t=0$:
\begin{align}
    \bar{\omega}&=\int_{-\infty}^{+\infty}\omega g(\omega)d\omega\label{eq:truthCalculation}\\
    &=\left. \frac{d\Phi(t)}{dt} \right| _{t=0}\nonumber,
\end{align}
and $t=0$ corresponds to the time when the $\pi/2$-pulse starts. This average frequency corresponds to the average field sensed by the probe weighted by the signal amplitude for the frequency interval.

The phase function $\Phi(t)$ can be constructed using the Hilbert transform. The Hilbert transform ($\mathcal{H}$) of an arbitrary function $u(t)$ is defined as \cite{Zygmund:1988}:
\begin{align}
    \mathcal{H}\{u(t)\}=\frac{1}{\pi}\lim_{\epsilon \rightarrow 0}\int_{\epsilon}^{+\infty}\frac{u(t+\tau)-u(t-\tau)}{\tau}d\tau.\label{eq:hilbert_transform}
\end{align}
Particularly, the Hilbert transform of $\exp(-t/T_{2})\cos(\omega t)$ ($\omega>0,t>0$) is $\exp(-t/T_{2})\sin(\omega t)$. According to Eq.~\ref{eq:fid_nonuniform}, the physical FID signal $f_{r}(t)$ is essentially a linear superposition of functions $\exp(-t/T_{2})\cos(\omega t+\phi_{0})$ with weight $Ng(\omega)$. Because the Hilbert transform is linear, the Hilbert transform, $f_{i}(t)$, of the FID signal must be the superposition of the $\exp(-t/T_{2})\sin(\omega t+\phi_{0})$ with the same weight. Therefore, 
\begin{align}
    f_{i}(t)&=N\exp\left(-\frac{t}{T_{2}}\right)\int_{-\infty}^{+\infty}g(\omega)\sin(\omega t + \phi_0)d\omega,\label{eq:fid_imaginary}\\
    &=A(t)\sin(\Phi(t)),\nonumber\\
    &={\text Im}(f(t)).\nonumber
\end{align}
Then the envelope function, $A(t)$, and the phase function, $\Phi(t)$, of an FID can be obtained by 
\begin{align}
    A(t)&=\sqrt{f_r^2(t)+f_i^2(t)},\label{eq:envelope}\\
    \Phi(t)&=\tan^{-1}(f_i(t)/f_r(t)).\label{eq:phase_hilbert}
\end{align}
The Hilbert transform can be performed via the Fourier transform ($\mathcal{F}$):
\begin{align}
    \mathcal{H}\{u(t)\}=\mathcal{F}^{-1}\{-i\text{sgn}(\omega)\mathcal{F}\{u(t)\}\}, \label{eq:hilbert_transform_fourier}
\end{align}
and therefore Fourier transform algorithms are often used to compute the Hilbert transform of a function.
Because the FID waveforms in this analysis are discrete, in this paper $\mathcal{H}$ and $\mathcal{F}$ represent discrete Hilbert and Fourier transforms.

The constant initial phase $\phi_0$ in Eq.~\ref{eq:fid_nonuniform} can be factored out, and thus, $\Phi(t)-\phi_0$ can be written explicitly as
\begin{align}
    \Phi(t)-\phi_0=\tan^{-1}\left(\frac{\int_{-\infty}^{+\infty}g(\omega)\sin(\omega t)d\omega}{\int_{-\infty}^{+\infty}g(\omega)\cos(\omega t)d\omega}\right).\label{eq:phase_explicit}
\end{align}
Therefore, $\Phi(t)-\phi_0$ is an odd function of $t$, and its Taylor expansion at $t=0$ contains only odd orders. The third and higher order derivatives of $\Phi(t)$ at $t=0$ are related to higher-order moments of $g(\omega)$ \cite{Cowan_1996}. 
The phase function is then fit to a truncated power series
\begin{align}
\Phi_{\text{fit}}(t)=\phi_{0}+p_{1}t+p_{3}t^{3}+p_{5}t^{5}+\cdots,\label{eq:fit_function}
\end{align}
and $\bar{\omega}$ is the fitted value of $p_{1}$ according to Eq.~\ref{eq:truthCalculation}.

\begin{figure}[h!]
\centering
\includegraphics[width=0.5\linewidth]{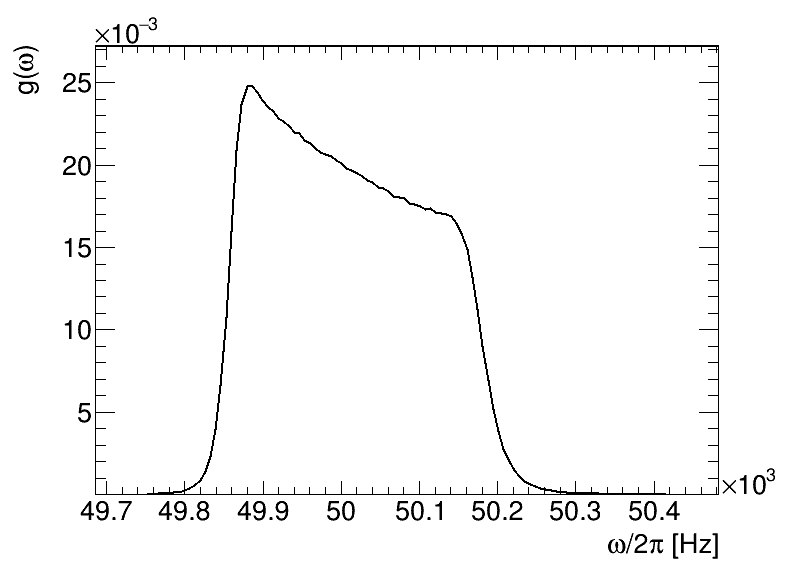}
\caption{Spectrum density function $g(\omega)$ for a simulated FID of an NMR probe that measures a magnetic field with both a first-order derivative (0.3~ppm/mm) and a second-order derivative (5~ppb/mm$^{2}$) along the probe axial direction. \label{fig:LineShape}}
\end{figure}

\begin{figure}[h!]
\centering
\subfloat[][FID  and its envelope] {\includegraphics[width=0.49\linewidth]{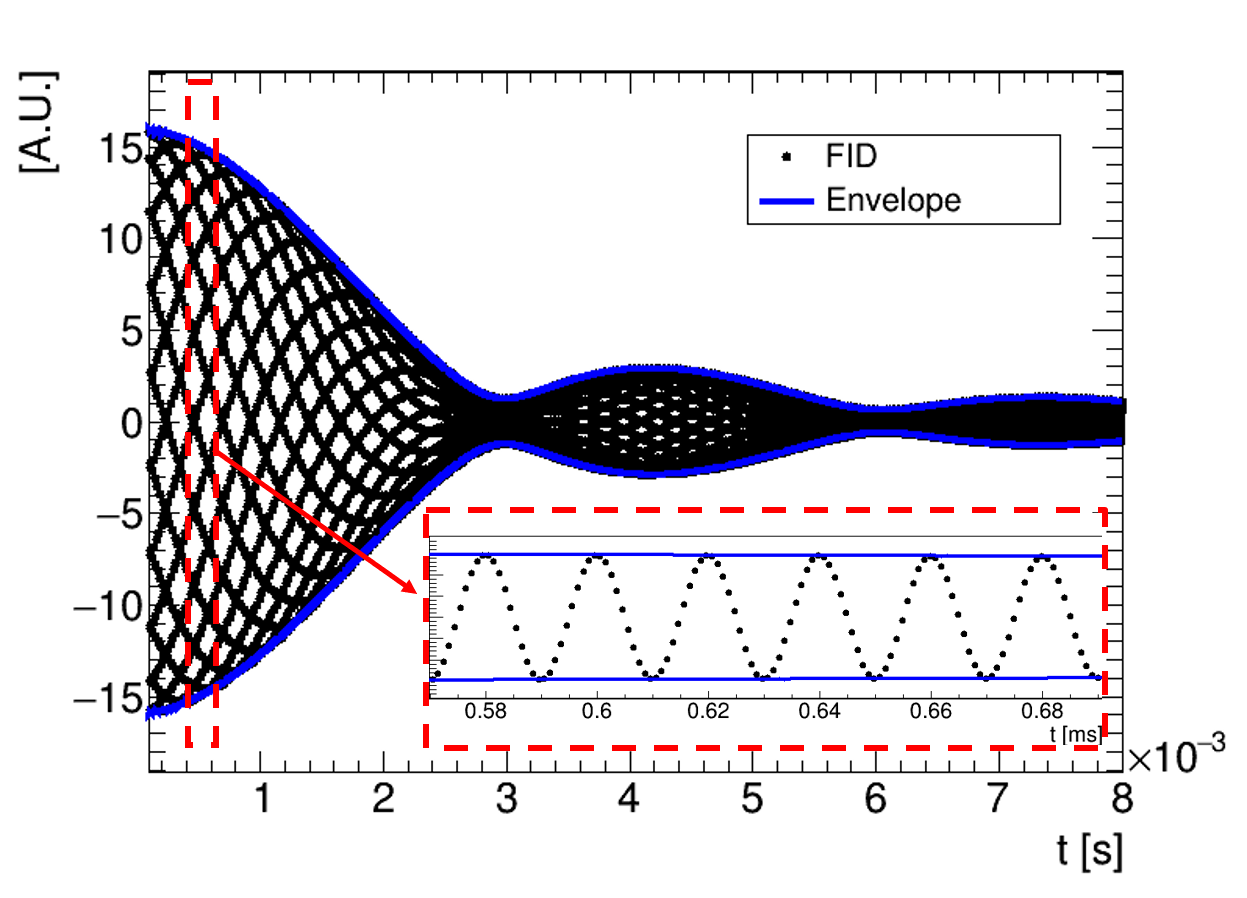} \label{fig:Fid}}
\subfloat[][Phase]{\includegraphics[width=0.49\linewidth]{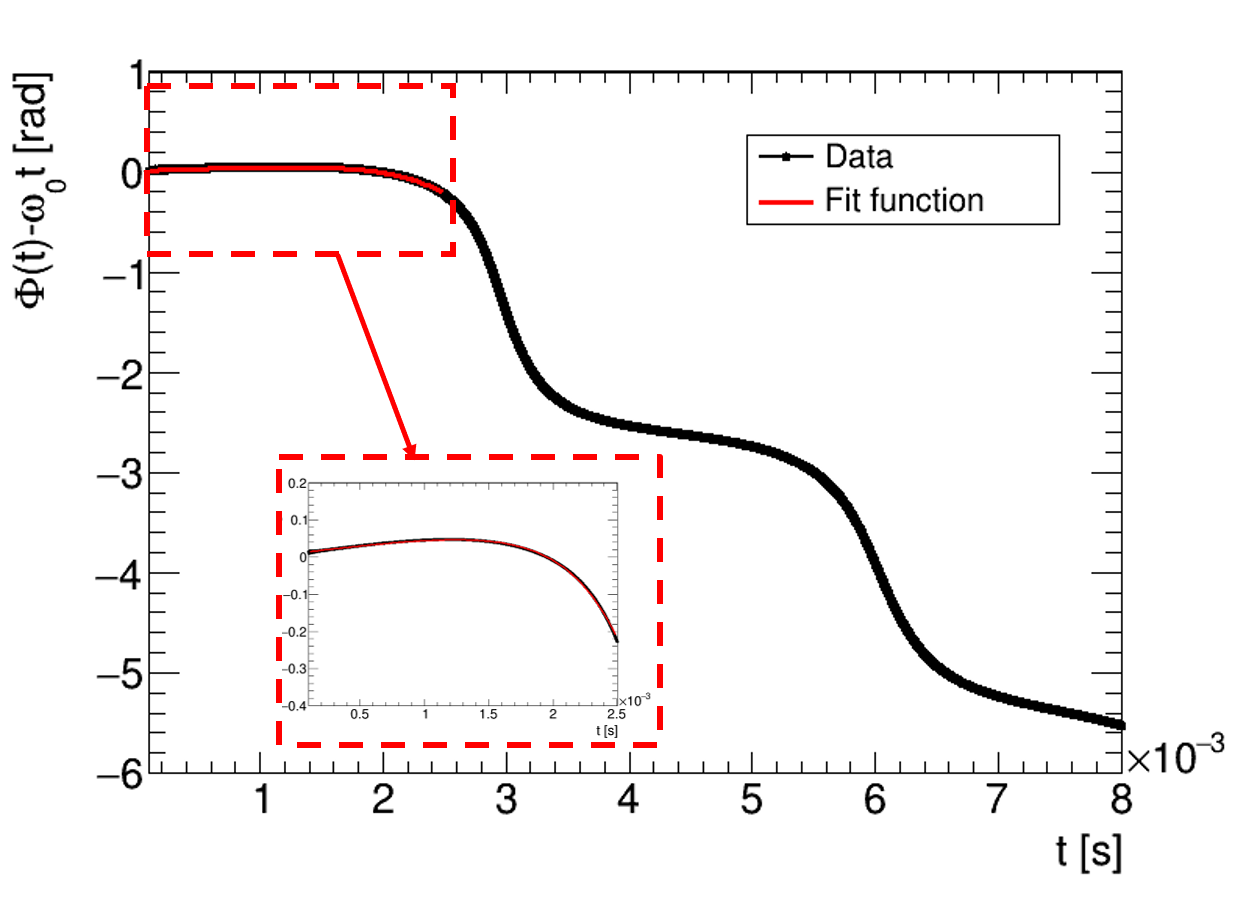}\label{fig:Phase}}
\caption{FID, envelope and phase. The pattern seen in the FID plot is an artifact due to the discretized data points. The insert in Fig.~\ref{fig:Fid} is a magnified view of the FID near $t=0.6$~ms to show its sinusoidal-oscillation pattern. In Fig.~\ref{fig:Phase}, to better visualize the non-linear component of the phase function, $\omega_0 t$ is subtracted from $\Phi(t)$, where $\omega_0$ is an angular frequency close to $\bar{\omega}$. The fit region is magnified.}
\label{fig:env_phi}
\end{figure}

The validity of this method has been studied with simulated FIDs that are constructed using artificial $g(\omega)$ functions. In this study, the $g(\omega)$ function is derived from a realistic magnetic field map in the Muon $g-2$ experiment \cite{hong2019magnetic} and the geometry of the NMR probes used in this experiment. 
The magnetic field in the muon beam storage ring is $\sim$1.45~T. The NMR probes for scanning and monitoring the magnetic field have a coil with a length of 1.5~cm and a diameter of 4.6~mm, and the detection material is petroleum jelly filled in a cylindrical cell inside the coil and that extends twice as long as the coil length. The proton-precession frequency in this magnetic field is about 61.79~MHz, and 
the local oscillator \cite{SG380} reference frequency is set to 61.74~MHz so that the frequency of the measured FID is near 50~kHz. The magnetic field has a peak-to-peak 90~ppm fluctuation around its $\sim$45-m perimeter. The fluctuations are short-ranged, resulting in gradients larger than 1~ppm/mm ($\sim$62$\times 2\pi$~Hz/mm in terms of angular frequency) at many locations. To exemplify the FID frequency extraction, an FID measured in a typical magnetic field with a gradient of 0.3~ppm/mm and a second-order derivative of 5~ppb/mm$^{2}$ along the probe axis is simulated, and the simulated spectrum density function is shown in Fig.~\ref{fig:LineShape}. Due to the nonzero second-order spatial derivative of the field, $g(\omega)$ is not symmetric and thus $\Phi(t)$ is nonlinear \cite{Cowan_1996}. The FID constructed using this $g(\omega)$ function is shown in Fig.~\ref{fig:Fid} together with the extracted envelope function. The extracted phase function is shown in Fig.~\ref{fig:Phase}, along with a fit to Eq.~\ref{eq:fit_function} truncated at the order of $t^{7}$ in the window of 0 to 2.5~ms. The fitted value of $p_{1}$ is different from the true value of $\bar{\omega}$ (evaluated using Eq.~\ref{eq:truthCalculation}) by 0.1$\times 2\pi$~Hz, well below the uncertainty budget of 0.6$\times 2\pi$~Hz \cite{Grange:2015fou} for the FID frequency extraction in the Muon $g-2$ experiment. The fit accuracy can be improved by adjusting the fit region and the truncation order of the fit function. For example, if the end of the fit range is reduced to the time when the FID envelope drops to 70\% of its maximum amplitude, and the truncation order is $t^{5}$, the difference between the fitted value and the truth of $\bar{\omega}$ is below 0.01$\times 2\pi$~Hz. In the following studies, this choice of fit range and truncation order is used.

The accuracy of the fit described above is achieved for an FID without noise or artifacts, even with a $\sim$350$\times 2\pi$~Hz full-width-half-maximum (FWHM) of the corresponding $g(\omega)$. As long as the fit range is within the Taylor series convergence radius of $\Phi(t)$, the fit accuracy can be improved by increasing the truncation order. However, the fit accuracy is also limited by the effects of artifacts, which will be described in Sec.~\ref{sec:artifact_uncertainty}.

%% file: Artifact.tex
\section{Artifacts and Systematic Uncertainties}
\label{sec:signal_artifacts}

The FID frequency extraction method described in Sec.~\ref{sec:method} relies on the fitting of the phase function, so it is crucial to understand how the artifacts, created by the discrete Hilbert-transform or intrinsic to the FID waveform, affect the phase function extraction. These artifacts and their effect in the phase function are discussed in Sec.~\ref{sec::discrete_hilbert} to Sec.~\ref{sec:artifact_phase}, and a mitigation method will be described in Sec.~\ref{sec:artifact_uncertainty}.

\subsection{Discrete Hilbert Transform of a Finite-length Waveform}
\label{sec::discrete_hilbert}

\begin{figure}[h]
\centering
\includegraphics[width=0.7\linewidth]{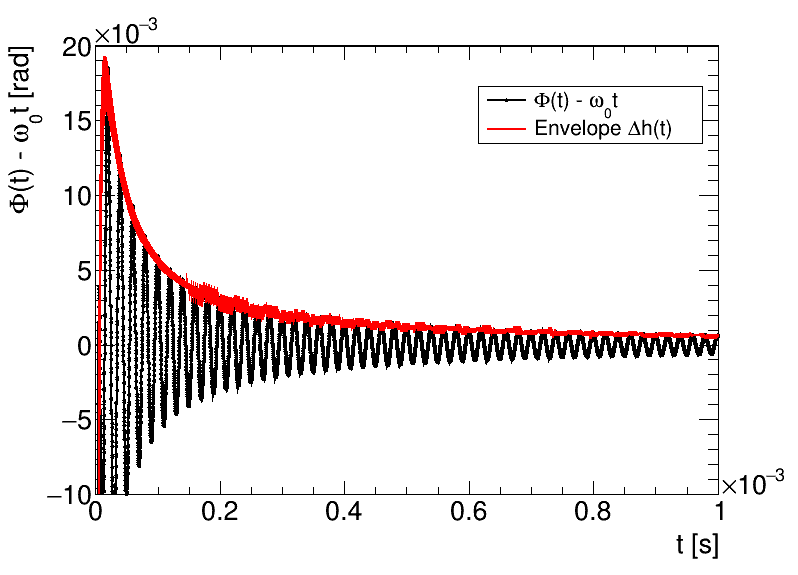}
\caption{The artifact $\Phi(t)-\omega_{0}t$ in the extracted phase function caused by the discrete Hilbert transform. Only the beginning section up to 1~ms is shown.}
\label{fig:phase_ripple_discrete_hilbert}
\end{figure}

The discrete Hilbert transform of the digitized FID waveform with a finite length does not produce the exact Hilbert transform for a continuous and infinitely-long function, and thus Eq.~\ref{eq:fid_imaginary} is not accurately produced. This artifact is obvious in the frequency domain. For the function $\cos(\omega_{0}t)$ with $\omega_{0}>0$, according to Eq.~\ref{eq:hilbert_transform_fourier}, the discrete Fourier transform (for $\omega \ge 0$) of its Hilbert transform is
\begin{align}
    \mathcal{F}\{\mathcal{H}\{\cos(\omega_{0}t)\}\}&=\frac{-i}{2}\sum_{k=0}^{T/\Delta t -1}\left(e^{i(\omega_{0}-\omega)k\Delta t}+e^{-i(\omega_{0}+\omega)k\Delta t}\right)\Delta t\label{eq:discrete_fourier_hilbert_cos}\\
    &=\frac{\Delta t}{2i}\frac{\sin(\frac{\omega_{0}-\omega}{2}T)}{\sin(\frac{\omega_{0}-\omega}{2}\Delta t)}e^{\frac{i(\omega_{0}-\omega)}{2}(T-\Delta t)}+\frac{\Delta t}{2i}\frac{\sin(\frac{\omega_{0}+\omega}{2}T)}{\sin(\frac{\omega_{0}+\omega}{2}\Delta t)}e^{\frac{-i(\omega_{0}+\omega)}{2}(T-\Delta t)},\nonumber
\end{align}
where $\Delta t$ is the sampling period and $T$ is the length of the digitized waveform. However, the discrete Fourier transform (for $\omega \ge 0$) of $\sin(\omega_{0} t)$, which is the exact Hilbert transform of $\cos(\omega_{0} t)$, is
\begin{align}
    \mathcal{F}\{\sin(\omega_{0}t)\}&=\frac{1}{2i}\sum_{k=0}^{T/\Delta t -1}\left(e^{i(\omega_{0}-\omega)k\Delta t}-e^{-i(\omega_{0}+\omega)k\Delta t}\right)\Delta t\label{eq:discrete_fourier_sin}\\
    &=\frac{\Delta t}{2i}\frac{\sin(\frac{\omega_{0}-\omega}{2}T)}{\sin(\frac{\omega_{0}-\omega}{2}\Delta t)}e^{\frac{i(\omega_{0}-\omega)}{2}(T-\Delta t)}-\frac{\Delta t}{2i}\frac{\sin(\frac{\omega_{0}+\omega}{2}T)}{\sin(\frac{\omega_{0}+\omega}{2}\Delta t)}e^{\frac{-i(\omega_{0}+\omega)}{2}(T-\Delta t)},\nonumber
\end{align}
whose second term in the final line is the negative of that in Eq.~\ref{eq:discrete_fourier_hilbert_cos}.
Comparing Eq.~\ref{eq:discrete_fourier_hilbert_cos} and Eq.~\ref{eq:discrete_fourier_sin} and those corresponding expressions for $\omega < 0$, the difference between the discrete Hilbert transform and the exact Hilbert transform of $\cos(\omega_{0}t)$ is
\begin{align}
  \Delta h(t) &:= \mathcal{H}\{\cos(\omega_{0}t)\}-\sin(\omega_{0}t) \label{eq:discrete_hilbert_artifact}\\
  &=\mathcal{F}^{-1}\left\{\frac{\text{sgn}(\omega)\Delta t}{i}\frac{\sin(\frac{\omega_{0}+\text{sgn}(\omega)\omega}{2}T)}{\sin(\frac{\omega_{0}+\text{sgn}(\omega)\omega}{2}\Delta t)}e^{\frac{-i\text{sgn}(\omega)\omega_{0}-i\omega}{2}(T-\Delta t)}\right\}.\nonumber
\end{align}
In the following example, $\Delta h(t)$ is computed numerically with $\omega_{0}=2\pi\times 50$~kHz, $\Delta t = 0.1$~ms, and $T=20$~ms. The value of $|\Delta h(t)|$ is large near the edges of the waveform as shown in Fig.~\ref{fig:phase_ripple_discrete_hilbert}, but if $t$ is two or more oscillation periods away from the edges, $|\Delta h(t)|$ is less than 1.5\% of the amplitude of the original waveform (which is 1 in this example) and $\Delta h(t)$ is a slow-varying function. With a non-zero $\Delta h(t)$, for $f(t)=\cos(\omega_{0}t)$, the extracted phase function is 
\begin{align}
    \Phi(t)&=\tan^{-1}\left(\frac{\sin(\omega_{0}t)+\Delta h(t)}{\cos(\omega_{0}t)}\right)\label{eq:phase_discrete_hilbert}\\
    &=\omega_{0}t+\cos(\omega_{0}t)\Delta h(t)-\cos(\omega_{0}t)\sin(\omega_{0}t)\Delta h^{2}(t)+\cdots.\nonumber
\end{align}
Therefore, $\Delta h(t)$ causes an artifact in the phase function $\Phi(t)$, which includes all terms on the right-hand side of Eq.~\ref{eq:phase_discrete_hilbert} except $\omega_{0} t$. Up to the linear order of $\Delta h(t)$, the artifact is an oscillation with an angular frequency $\omega_{0}$ and an envelope $\Delta h(t)$ as shown in Fig. 3. The method of mitigating this artifact is described in Sec.~\ref{sec:artifact_uncertainty}.

\subsection{Artifacts of the FID Waveform}
\label{sec:artifact_waveform}

\begin{figure}[h]
\centering
\includegraphics[width=0.7\linewidth]{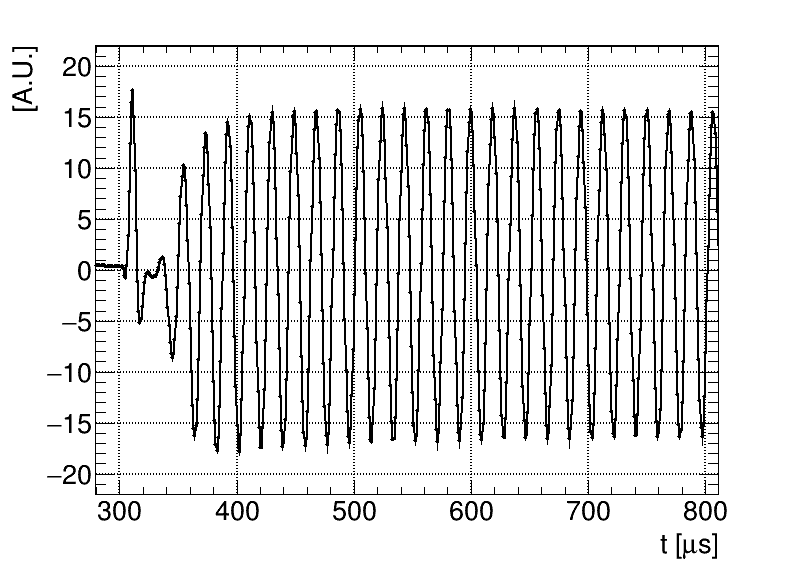}
\caption{The early section of a trolley FID exhibits the signal distortion and the time-dependent baseline.}
\label{fig:trolley_fid}
\end{figure}

\begin{figure}[h]
\centering
\includegraphics[width=0.7\linewidth]{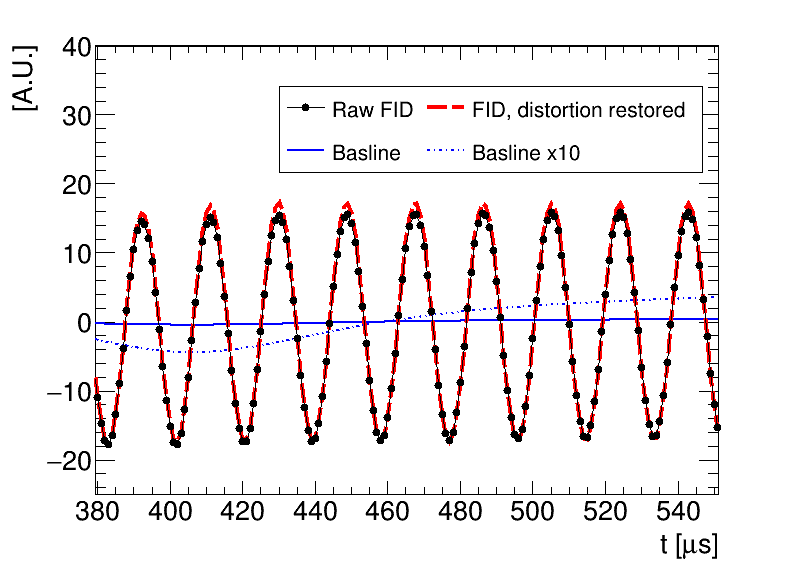}
\caption{The extracted baseline and FID with the positive amplitude corrected. To better visualize the shape of the extracted baseline, the 10-times exaggerated baseline is shown as the dashed blue line.}
\label{fig:baseline_distortion}
\end{figure}

In the following parts of this section, FIDs from the magnetic field scanner probe \cite{Corrodi_2020} in the Muon $g-2$ Experiment are chosen for illustration and algorithm validation. One example FID is shown in Fig.~\ref{fig:trolley_fid}. The $\pi/2$-pulse is fired at 300~$\mu$s, and the signal amplifier is turned on at 350~$\mu$s. It is obvious that the upper and lower envelopes do not have the same shape before $\sim$600~$\mu$s, indicating a time-dependent baseline or signal distortion. By definition, a baseline is a slow-varying function added to the ideal FID. 
Therefore, the baseline of a measured FID waveform can be determined by finding the line that intersects with the FID waveform at even intervals within the range of one or two complete oscillations, assuming the phase function is linear in this time range. 
The extracted baseline for the FID in Fig.~\ref{fig:trolley_fid} is shown in Fig.~\ref{fig:baseline_distortion}. For this FID, the maximum of the baseline absolute value is $<$0.5\% of the amplitude of the FID.

\begin{figure}[h]
\centering
\includegraphics[width=0.7\linewidth]{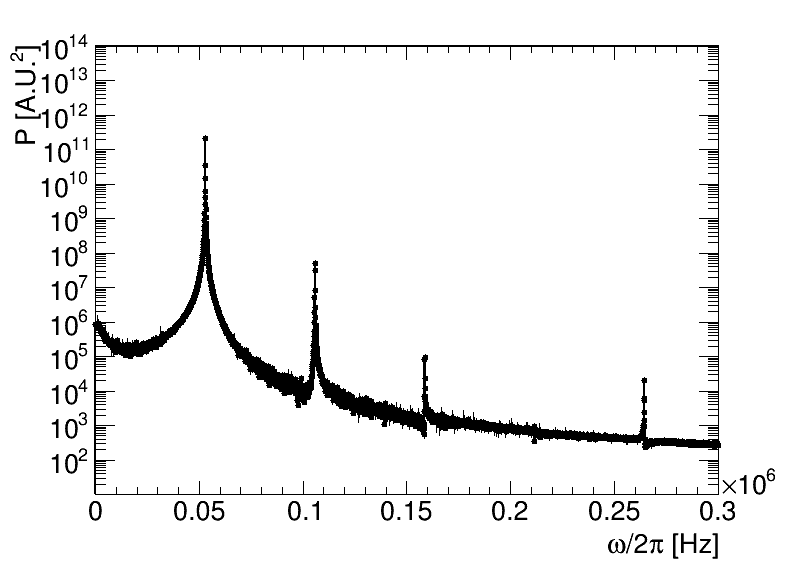}
\caption{FID power spectrum with higher harmonics. The power is defined as the square of the signal.}
\label{fig:trolley_power_spectrum}
\end{figure}

After the baseline is determined, the positive amplitude (from the baseline to a local maximum) and the negative amplitude (from the local minimum to the baseline) of the FID are investigated. Throughout the entire FID, the positive amplitude is consistently smaller than the negative amplitude. This effect is also illustrated in Fig.~\ref{fig:baseline_distortion}, and in this beginning part of the FID, the positive amplitude is $\sim$10\% smaller than the negative amplitude. 
In the frequency domain, such a waveform distortion results in higher-order harmonics in the power-density spectrum as shown in Fig.~\ref{fig:trolley_power_spectrum}


The time-dependent baseline and the waveform distortion are caused by the readout electronics, and they can be reduced by optimizing the circuit design. However, they may be irreducible when there are constraints on the choice of components, for example, power consumption, vacuum compatibility, and magnetic footprint. In these cases, it is important to understand how these artifacts affect the extracted phase function and how to mitigate their effects.

\subsection{Effects of the Baseline and Signal Distortion on the Phase Function}
\label{sec:artifact_phase}


Because the difference between the FIDs with and without the artifacts is usually less than 10\% of the FID oscillation amplitude in its full range, the artifacts can be treated as small perturbations on the FID signal. In this section, the perturbations on the phase function are derived analytically up to the leading order. 

Suppose the measured FID waveform with a nontrivial baseline is $f_{r}(t)=A(t)\cos(\Phi(t))+b(t)$, where $b(t)$ is the baseline. The Hilbert transform of $f_{r}$ is $f_{i}(t)=A(t)\sin(\Phi(t))+b_{i}(t)$, where $b_{i}(t)$ is the Hilbert transform of $b(t)$. To simplify the following expressions, define $\alpha(t)=b(t)/A(t)$ and $\alpha_{i}(t)=b_{i}(t)/A(t)$. The envelope and phase of $f_{r}(t)$ can be extracted using Eq.~\ref{eq:envelope} and Eq.~\ref{eq:phase_hilbert}. Alternatively, one can also extract them by explicitly writing the complex function $f_{r}(t)+if_{i}(t)$ into the modulus-argument form while keeping $\alpha$ and $\alpha_{i}$ up to the linear order:

\begin{align}
    f_{r}+if_{i}&=A(\exp(i\Phi)+\alpha+i\alpha_{i})\label{eq:generic_fid_baseline}\\
    &=A\exp(i\Phi)(1+(\alpha+i\alpha_{i})\exp(-i\Phi))\nonumber\\
    &=A\exp(i\Phi)(1+\alpha\cos(\Phi)+\alpha_{i}\sin(\Phi)+i(\alpha_{i}\cos(\Phi)-\alpha\sin(\Phi))\nonumber\\
    &\approx A\exp(i\Phi)\sqrt{1+2\alpha\cos(\Phi)+2\alpha_{i}\sin(\Phi)}\nonumber\\
    &\times\exp\left(i\tan^{-1}\left(\frac{\alpha_{i}\cos(\Phi)-\alpha\sin(\Phi)}{1+\alpha\cos(\Phi)+\alpha_{i}\sin(\Phi)}\right)\right)\nonumber\\
    &\approx A(1+\alpha\cos(\Phi)+\alpha_{i}\sin(\Phi))\exp(i\Phi+i\alpha_{i}\cos(\Phi)-i\alpha\sin(\Phi)).\nonumber
\end{align}



Assuming the baseline is slow-varying compared to the fast oscillation $\cos(\Phi(t))$, $b(t)$ is approximately a constant and $b_{i}(t)$ is approximately zero. After dropping $\alpha_{i}$, the extracted FID envelope and phase become
\begin{align}
A_{\text{ext}}(t)&=A(t)+b(t)\cos(\Phi(t)),\\
\Phi_{\text{ext}}(t)&=\Phi(t)-\alpha(t)\sin(\Phi(t)).
\end{align}
Therefore, the baseline results in ripples $b(t)\cos(\Phi(t))$ in the envelope function, and also ripples $-\alpha(t)\sin(\Phi(t))$ in the phase function. The frequencies of the envelope ripple and the phase ripple are the same as the FID frequency, but the phase of the ripple in the phase function is $\pm\pi/2$ different from the FID oscillation phase, where the $\pm$ sign depends on the sign of $\alpha(t)$. The amplitude of the ripple of $A(t)$ depends on the baseline size $b(t)$, while the amplitude of the phase ripple depends on the baseline-to-amplitude ratio $\alpha(t)$.

For the signal distortion, it is easier to treat them as higher-order harmonics. Suppose the $m$'th order harmonic term is $\beta(t)A(t)\exp(im\Phi(t))$. A complex FID waveform with this term is $f(t)=A(t)(\exp(i\Phi(t))+\beta(t)\exp(im\Phi(t)))$, and keeping up to the linear order of $\beta(t)$, it becomes 
\begin{align}
    f&=A\exp(i\Phi)(1+\beta\exp(i(m-1)\Phi))\label{eq:second_harmonic_effect_dev}\\
    &\approx A(1+\beta\cos((m-1)\Phi))\exp(i\Phi+i\beta\sin((m-1)\Phi)).\nonumber
\end{align}
Therefore, higher-order harmonics also result in ripples in the envelope and phase function. The ripple frequency of an $m$'th order harmonic term is $m-1$ times the FID base frequency. Particularly, the slow-varying baseline can be treated as the case when $m=0$, and the ripple frequencies for the baseline and the second-order harmonic term are the same, which is the FID base frequency.

\begin{figure}[h]
\centering
\includegraphics[width=0.7\linewidth]{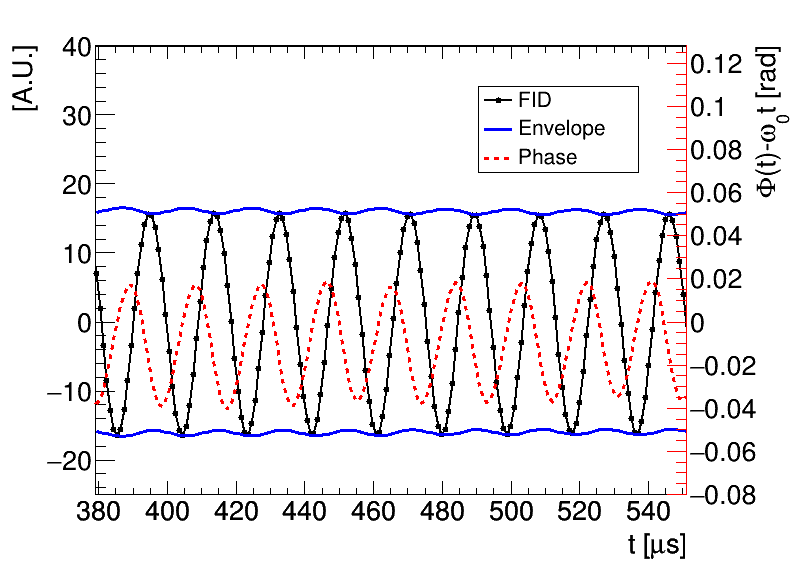}
\caption{Ripples on the extracted FID envelope (blue) and phase function (red). To better visualize the ripples on the phase function, $\omega_0 t$ is subtracted from $\Phi(t)$, where $\omega_0$ is an angular frequency close to $\bar{\omega}$.}
\label{fig:trolley_env_phase}
\end{figure}

The extracted envelope and phase functions of the FID in Fig.~\ref{fig:trolley_fid} are shown in Fig.~\ref{fig:trolley_env_phase}. For this FID, the second harmonic term $\beta(t)$ dominates the other harmonic terms and the baseline. The phases of the ripples in the extracted envelope and phase functions are consistent with the derivation described above.

\subsection{Artifact-related systematic uncertainty and mitigation method}
\label{sec:artifact_uncertainty}

The ripples caused by the artifacts in the phase function will affect the fit result of the average frequency. The bias of the fit result is sensitive to the starting and ending points of the fit range relative to the ripple phase. We simulated an FID with a $\sim$50~kHz frequency and with artifacts that made the amplitude of the phase ripple 0.03~rad. For such an FID, if the start of the fit range is fixed and the width of the fit range varies within $1\pm0.02$~ms, the bias of the $\bar{\omega}$ extraction caused by the phase ripple varies within $\pm0.6\times 2\pi$~Hz. The magnitude of the bias depends strongly on the overall fit range. The longer the fit range is, the smaller the bias is.   

Because the ripple in the phase function oscillates at the same frequency as the FID, it can be mitigated by a moving-average smoothing method with the averaging window $T_{w}$, which is the same as the FID oscillation period. $T_{w}$ can be determined using an approximated FID frequency found by fitting the extracted $\Phi(t)$ without the ripple mitigation. 
If the ripples are totally eliminated, the extracted FID frequency will not be sensitive to the end points of the fit region within an FID cycle. However, the smoothing is discrete and thus $T_{w}$ cannot perfectly match the FID cycle period $T_{0}$. If $\Delta T=T_{0}-T_{w}$ is small, the amplitude of the remaining ripple after smoothing is $\Delta T/T_{0}$ of the original amplitude. Moreover, the smoothing distorts the phase function for samples within $T_{w}$ from the edge of the FID. Because the discrete Hilbert transform also introduces large ripples near the edges, the actual fit window should start at least one or two oscillation cycles from the FID sample with the largest amplitude. If the smoothing is applied multiple times, then multiples of $T_{w}$ should be avoided when determining the fit range.

For those FIDs with a fast-decreasing envelope or a fast-varying baseline, $\alpha(t)$ varies significantly within one oscillation period and thus the smoothing is less effective. For such FIDs, the phase function $\Phi(t)$ usually has large nonlinear terms, so the systematic bias of the FID frequency extraction becomes significant. In these cases, it is better to use the simulated FID to estimate the systematic biases as long as the analysis algorithms and parameters (like the truncation order of the fit function) are chosen the same as those in real measurements.

There are other ways to mitigate the effects of the baseline and the signal distortion, but the running-average phase smoothing method has more advantages. Although the baseline can be extracted from the FID waveform as described in Sec.~\ref{sec:artifact_waveform} and then corrected, it is difficult to formulate the systematic uncertainty caused by an imperfect baseline extraction. The slow-varying baseline and higher-order harmonics can be filtered out in the frequency domain, but such filters also affect the phase function extraction and complicate the systematic uncertainty analysis. On the other hand, the running-average phase smoothing method is simple to implement, and the systematic uncertainty analysis described above is also straight-forward. The smoothing operation can also be easily incorporated in the statistical uncertainty analysis described in Sec.~\ref{sec:noise}.


%% file: Noise.tex
\section{Noise and Statistical Uncertainty}
\label{sec:noise}

The statistical uncertainty of $\bar{\omega}$ is given by the minimum-$\chi^{2}$ fit of Eq.~\ref{eq:fit_function} to the extracted phase function $\Phi(t)$, provided that the uncertainty of each $\Phi(t)$ sample and the correlation between samples are set correctly. In this section, the noise in $\Phi(t)$ is derived given the signal noise. The phase noise covariance matrix for constructing the $\chi^{2}$, bias of the fit results, and the goodness of the fit are investigated for the white noise and a few generic noise spectra. It is important to obtain the correct expression of the $\chi^{2}$ and make sure that the covariance matrix is invertible so that the fit yields unbiased and consistent results of $\bar{\omega}$ and its error bar. A method of handling non-invertible covariance matrices is described. The performances of two other less rigorous methods, the {\em unweighted} and {\em diagonal} minimum-$\chi^{2}$ fit methods, are discussed as well.

\subsection{Noise in the Phase Function}
\label{subsec:phase_noise_derivation}

The noise in the detected signal is a random sequence $N(t)$ added to the FID waveform: $f_r(t)=A(t)\cos(\Phi(t))+N(t)$. Following the same procedure described in Sec.~\ref{sec:artifact_phase} and replacing $b(t)$ with $N(t)$, one gets the complex form of the FID waveform with noise $N(t)$
\begin{align}
    f_{r}+if_{i}&=A(1+n\cos(\Phi)+n_{i}\sin(\Phi))\label{eq:fid_with_noise}\\
    &\times\exp(i\Phi+i n_{i}\cos(\Phi)-i n\sin(\Phi)),\nonumber
\end{align}
where $n(t)=N(t)/A(t)$ and $n_{i}(t)=\mathcal{H}\{N\}(t)/A(t)$. Therefore, the noise in the phase function is 
\begin{align}
    n_{\phi}(t)=n_{i}(t)\cos(\Phi(t))-n(t)\sin(\Phi(t)).\label{eq:phase_noise}
\end{align}
Unlike the slow-varying $b(t)$, the Hilbert transform of $N(t)$ is not negligible and must be kept in the noise analysis. This formula has been verified using simulated FIDs with injected noises.


\subsection{White Noise}
\label{subsec:white_noise}

For simplicity, we first assume that $N(t)$ is a Gaussian white noise, and we let the distribution of $N(t)$ have a mean of zero and a standard deviation of $\sigma_N$. The standard deviation of $n(t)$ thus increases as $A(t)$ decreases with time. For a white noise $N(t)$, different noise samples are statistically independent, so different samples of $n(t)$ are also independent. Because $N_{i}(t)$ is derived from $N(t)$, the correlation between samples of $N(t)$ and $N_{i}(t)$ must be taken into account. So is the correlation between samples of $n(t)$ and $n_{i}(t)$. If the Hilbert transform is performed via discrete Fourier transform as in Eq.~\ref{eq:hilbert_transform_fourier}, the covariance matrix element for sample-$j$ from $n(t)$ and sample-$k$ from $n_{i}(t)$ is then (see Appendix~\ref{app:correlation_n_ni})
\begin{align}
    \text{COV}(n(t_{j}),n_{i}(t_{k}))=\frac{1-(-1)^{k-j}}{\pi(k-j)}\frac{\sigma_{N}^{2}}{A(t_{j})A(t_{k})}\label{eq:cov_n_ni}
\end{align}
for $j \neq k$. For $j=k$, the matrix element is zero. Among the samples of $n_{i}(t)$ (see Appendix~\ref{app:correlation_ni_ni})
\begin{align}
    \text{COV}(n_{i}(t_{j}),n_{i}(t_{k})) \approx \frac{\sigma_{N}^{2}}{A^{2}(t_{j})}\delta_{jk}\label{eq:cov_ni_ni},
\end{align}
if the two samples are not close to the ends of the sequence \footnote{The accurate expression is derived in Appendix~\ref{app:correlation_ni_ni}. The approximation of the diagonal element of $\text{COV}(n_{i}(t_{j}),n_{i}(t_{j}))$ is about 2.5\% off from the true value for $j=7$.}. According to Eq.~\ref{eq:phase_noise}, \ref{eq:cov_n_ni} and \ref{eq:cov_ni_ni} the covariance matrix for $n_{\phi}(t)$ can be calculated:
\begin{align}
    \Sigma_{jk}&=\text{COV}(n_{\phi}(t_{j}),n_{\phi}(t_{k}))\label{eq:phase_noise_cov}\\
    &=\frac{\sigma_{N}^{2}}{A^{2}(t_{j})}\delta_{jk}+(1-\delta_{jk})\sigma_{N}^{2}\frac{1-(-1)^{k-j}}{\pi(k-j)}\left(\frac{\cos(\Phi(t_{j}))\sin(\Phi(t_{k}))}{A(t_{j})A(t_{k})}-\frac{\cos(\Phi(t_{k}))\sin(\Phi(t_{j}))}{A(t_{j})A(t_{k})}\right),\nonumber
\end{align}
where $\delta_{ij}$ is the Kronecker Delta. The covariance matrix in Eq.~\ref{eq:phase_noise_cov} parameterizes the statistical distribution of the $\Phi(t)$ fluctuations. Therefore, when fitting the phase function, the $\chi^{2}$ to be minimized is
\begin{align}
    \chi^{2}&=(\Phi(t_{j})-\Phi_{\text{fit}}(t_{j}))\Sigma_{jk}^{-1}(\Phi(t_{k})-\Phi_{\text{fit}}(t_{k}))\label{eq:chi2_with_cov}\\
    &=(\Phi-\Phi_{\text{fit}})^{T}\Sigma^{-1}(\Phi-\Phi_{\text{fit}}),\nonumber
\end{align}
where $\Phi_{\text{fit}}(t)$ is the polynomial fit function defined in Eq.~\ref{eq:fit_function}. The standard minimum-$\chi^{2}$ fit procedure then yields the fit value of $\bar{\omega}$ and its statistical uncertainty $\sigma_{\omega}$.

\begin{figure}[h!]
\centering
\includegraphics[width=0.7\linewidth]{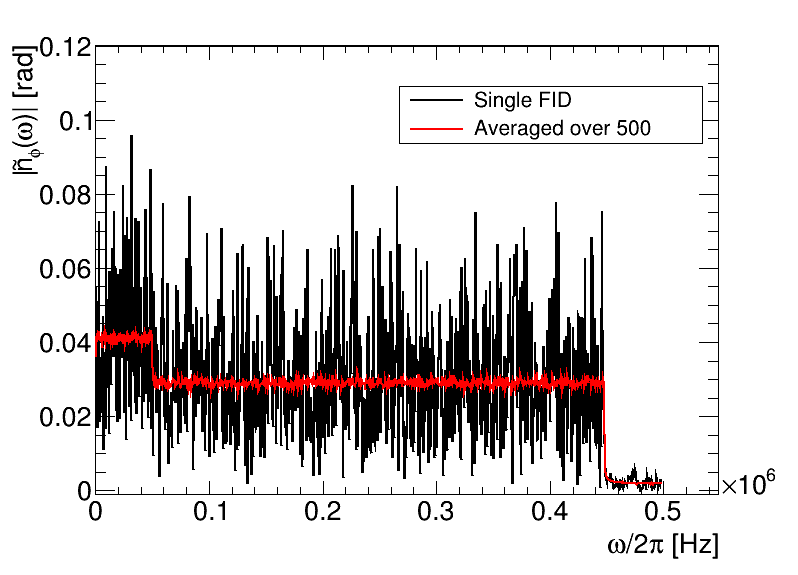}
\caption{Spectrum of the amplitude of $\tilde{n}_{\phi}(\omega)$. The noise has a standard deviation that equals 2\% of the FID maximum amplitude, and the average FID frequency is 50~kHz. The black curve is for one instance of the noise, and the red curve is obtained by averaging over 500 simulated FIDs with the same signal but independent noises. \label{fig:PhaseNoiseAmpSpectrum}}
\end{figure}

However, the matrix $\Sigma$ is nearly singular and becomes difficult to invert numerically. The approximate singularity of $\Sigma$ indicates that there are strong constraints on the $n_{\phi}(t)$ elements. This can be better revealed in the frequency domain. For a typical FID with a slow-varying envelope and a nearly linear phase, assuming $A(t)$ is a constant and $\Phi(t)=\omega_{0}t$, the Fourier Transform of $n_{\phi}$ is 
\begin{align}
    \tilde{n}_{\phi}(\omega)=\frac{i}{2}(\tilde{n}(\omega-\omega_{0})(1-\text{sgn}(\omega-\omega_{0}))-\tilde{n}(\omega+\omega_{0})(1+\text{sgn}(\omega+\omega_{0}))).\label{eq:fourier_n_phi}
\end{align}
Because the Fourier Transform is discrete, $\omega$ in Eq.~\ref{eq:fourier_n_phi} ranges from $-\pi/\Delta t$ to $+\pi/\Delta t$. The amplitude spectrum of $\tilde{n}_{\phi}(\omega)$ for such a typical FID is shown in Fig.~\ref{fig:PhaseNoiseAmpSpectrum}. Since $n_{\phi}(t)$ is a real function, $\tilde{n}_{\phi}(-\omega)=\tilde{n}_{\phi}(\omega)$ and the following discussions are for $\omega \ge 0$.

For $\omega<\omega_{0}$, the amplitude of $\tilde{n}_{\phi}(\omega)$ is $\sqrt{2}$ times that for $\omega_{0}<\omega<\pi/\Delta t-\omega_{0}$, because 
\begin{align}
    \tilde{n}_{\phi}(\omega)=i(\tilde{n}(\omega_0-\omega)-\tilde{n}(\omega+\omega_{0})),
\end{align}
which is a linear combination of two independent frequency components. If the independent variable $\omega$ of $\tilde{n}(\omega)$ is greater than the Nyquist angular frequency $\pi/\Delta t$, $\tilde{n}(\omega)$ is close to zero. Therefore, for $\omega>\pi/\Delta t-\omega_0$, $\tilde{n}_{\phi}(\omega)=-i\tilde{n}(\omega+\omega_{0})\approx 0$. After expressing $\tilde{n}_{\phi}(\omega)$ explicitly in terms of $n_{\phi}(t)$, one gets $\omega_{0}T/2\pi$ constraint equations for $\omega>\pi/\Delta t-\omega_{0}$:

\begin{align}
    \Sigma_{j}n_{\phi}(j\Delta t)e^{-i\omega(j\Delta t)}=0.
    \label{eq:spectrum_constraint}
\end{align}
Therefore, the degrees of freedom for $n_{\phi}(t)$ is reduced by $\omega_{0}T/2\pi$, which makes $\Sigma$ singular. A more detailed explanation is given in Appendix~\ref{app:rank_covariance_matrix}. The most straightforward way to remove these almost-redundant degrees of freedom in $n_{\phi}(t)$ is to down-sample 
$n_{\phi}(t)$ by a factor of two before fitting so that $n_{\phi}(t)$ does not have Fourier components at those high angular frequencies. Moreover, according to Eq.~\ref{eq:phase_noise_cov}, the off-diagonal elements are zero if $k-j$ is a even number. In this instance, the covariance matrix $\Sigma$ for the down-sampled $n_{\phi}(t)$ is a diagonal matrix
\begin{align}
    \Sigma_{jk}=\frac{\sigma_{N}^{2}}{A^{2}(t_{2j})}\delta_{jk},
\end{align}
which simplifies the computation of its inverse.

\begin{figure}[h!]
\centering
\includegraphics[width=0.7\linewidth]{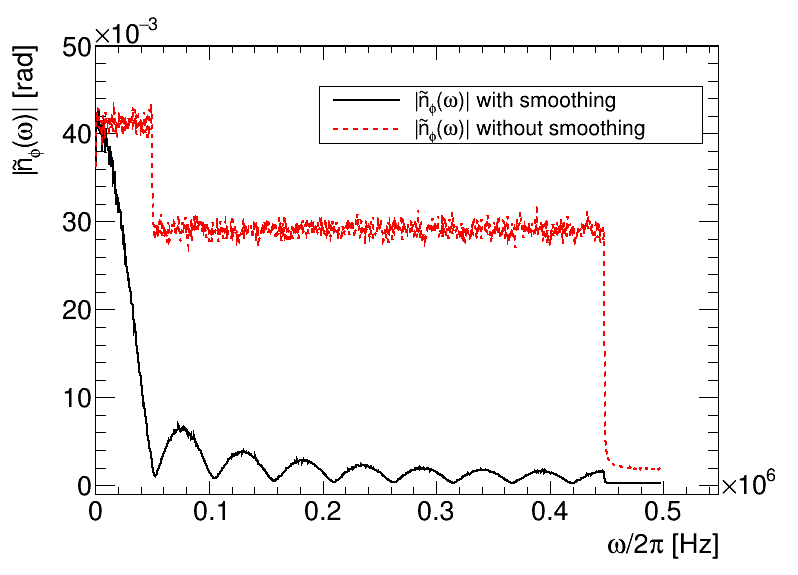}
\caption{Spectrum of the amplitude of $\tilde{n}_{\phi}(\omega)$ for smoothed $n_{\phi}(t)$, compared with that for the phase noise without smoothing as shown in Fig.~\ref{fig:PhaseNoiseAmpSpectrum}. \label{fig:PhaseNoiseAmpSpectrumSmooth}}
\end{figure}

The smoothing method described in Sec.~\ref{sec:artifact_uncertainty} for artifact mitigation affects the covariance matrix $\Sigma_{jk}$ as well. The smoothing can be expressed in a matrix form as
\begin{align}
    \Phi_{S}(t_{j})=S_{jk}\Phi(t_{k}),
\end{align}
and for $t$ far from the ends of the sequence (more than $W/2$ from each end)
\begin{align}
    S_{jk}=\frac{1}{W+1}\text{ for }\left|j-k\right|\le W/2,
\end{align}
where $W$ is the smoothing window size. The covariance matrix for the smoothed phase function is then $S\Sigma S^{T}$. From another point of view, the smoothing operation is a convolution of $n_{\phi}(t)$ with a square-pulse kernel function, and thus, in the frequency domain, the Fourier transform of $n_{\phi}(t)$ is multiplied with the Fourier transform of the square-pulse kernel function, which is a sinc function $\sin(\pi T_{w}\omega)/(\pi T_{w}\omega)$ with $T_{w}$ representing the duration of the smoothing window. After the smoothing, the noise spectrum becomes the black curve shown in Fig.~\ref{fig:PhaseNoiseAmpSpectrumSmooth}. Therefore, the smoothing operation is a low-pass filter with zeros at frequencies of multiples of $1/T_{w}$ that greatly suppresses frequencies higher than $1/T_{w}$. Applying the smoothing function multiple times will further suppress high-frequency noise components. As discussed above, to make the covariance matrix of the smoothed phase noise regular, the phase function has to be down-sampled so that $\tilde{n}_{\phi}(\omega) \neq 0$ up to the Nyquist angular frequency after the down-sampling. For a single-iteration smoothing, the down-sample factor should be at least $T_{w}/(2\Delta t)$.

The scheme of obtaining the covariance matrix described above was verified using simulated FIDs with the same signal and 500 independent white-noise waveforms. The bias and consistency of the extracted $\bar{\omega}$ and its statistical uncertainty are also investigated in this way. Fitting each of these FIDs yields $\bar{\omega}$, $\sigma_{\omega}$ and $\chi^{2}/\nu$, where $\nu$ is the degree of freedom. The mean of the extracted $\bar{\omega}$ is statistically consistent with the true value used in the simulation, and the standard deviation of these 500 $\bar{\omega}$ values is statistically consistent with the mean of the 500 $\sigma_{\omega}$ values. The distribution of $\chi^{2}/\nu$ is centered around 1. This test was performed for FIDs with different $T_{2}^{*}$ (the time when the envelope first decays to $1/e$ of the initial FID amplitude) values and phase non-linearities, and this fit scheme always yielded error bars consistent with the statistics and $\chi^{2}/\nu$ consistent with 1. Because the fit yields a $\chi^{2}/\nu$ consistent with 1, the goodness of the fit can be tested using a $\chi^{2}$-test. Then, one can use the goodness of the fit to determine whether the truncation order of the fit function is sufficient. 

\begin{figure}[h!]
\centering
\includegraphics[width=0.7\linewidth]{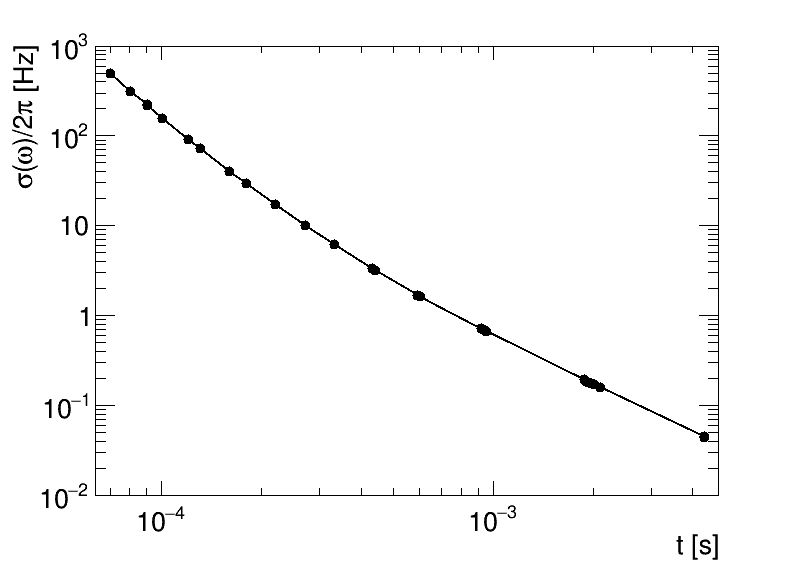}
\caption{$\sigma_{\omega}$ versus the actual length of the fit window. Truncation order is set to $t^{5}$ and $\sigma_{N}$ is 0.16\% of the maximum amplitude of the FID. \label{fig:ResolutionVsLength}}
\end{figure}

The statistical uncertainty ($\sigma_{\omega}$) of the extracted average frequency $\bar{\omega}$ increases with the noise-to-signal ratio, and decreases with the length of the fit window. It also increases with the truncation order of the fit polynomial due to the increase of degrees of freedom. The fit window and truncation order can be optimized in order to minimize the total uncertainty depending on how non-linear the phase function is. In principle, $\sigma_{\omega}$ also depends on the shape of the envelope function $A(t)$. To study this effect, we determined the $\sigma_{\omega}$ for simulated FIDs with different $T_{2}^{*}$ values and envelope shapes. The fit window is adjusted accordingly as described in Sec.~\ref{sec:method}. To generate such set of FIDs, one can scan through various ranges of first and second order spatial derivatives of the magnetic field where the probe is placed. As shown in Fig.~\ref{fig:ResolutionVsLength}, the relationship between $\sigma_{\omega}$ and the actual fit window length has a low dispersion, indicating that under the influence of the same noise, $\sigma_{\omega}$ depends predominantly on the fit window length, not the shape of $A(t)$. 

\subsection{Generic Noise Spectrum}
\label{subsec:general_noise}

\begin{figure}[h!]
\centering
\subfloat[][$\tilde{N}(\omega)$] {\includegraphics[width=0.49\linewidth]{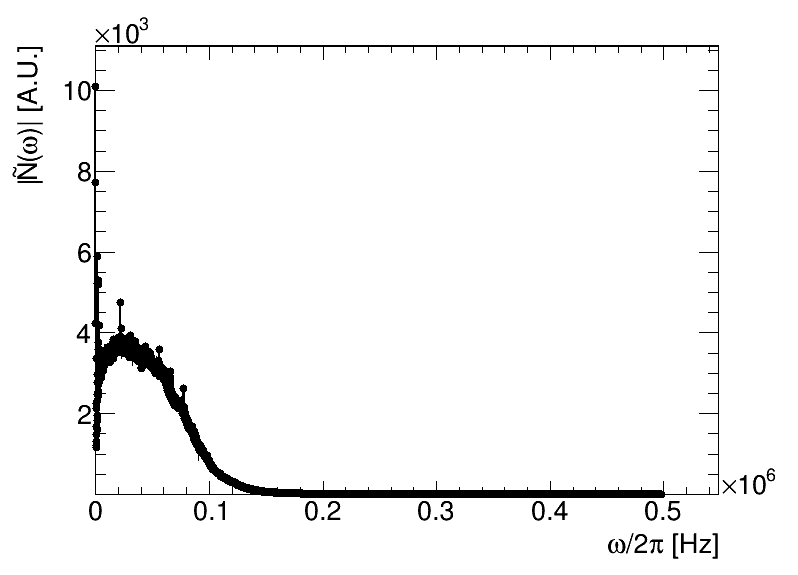} \label{fig:TrolleyNoiseFreqAmp}}
\subfloat[][$\tilde{n}_{\phi}(\omega)$] {\includegraphics[width=0.49\linewidth]{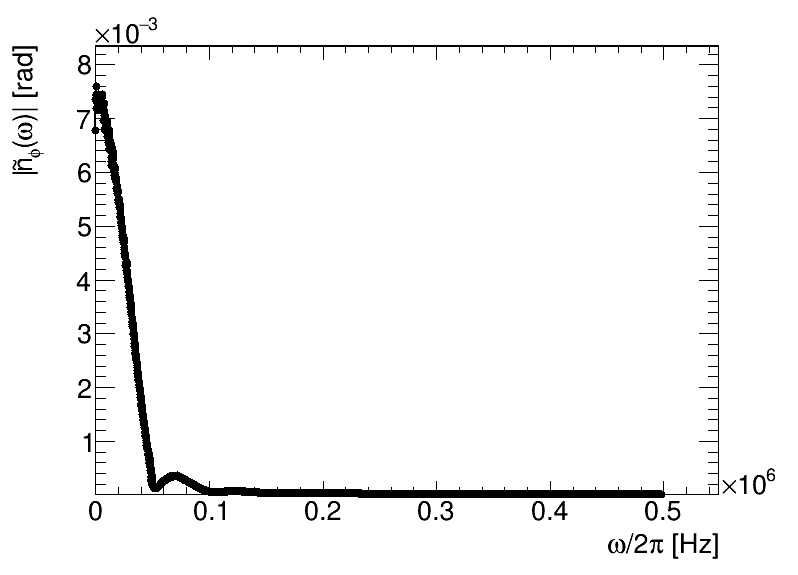} \label{fig:TrolleyPhaseNoiseFreqAmp}}
\caption{Frequency-domain spectra for noises in the signal ($\tilde{N}(\omega)$) and the smoothed phase function ($\tilde{n}_{\phi}(\omega)$) of the Muon $g-2$ magnetic field scanner probe. \label{fig:TrolleyNoiseAmpSpectrum}}
\end{figure}

The white noise model is a good approximation of noises in a wide range of magnetometer signals. In some applications, low-pass or band-pass filters are used to improve the signal-to-noise ratio of the FID. For example, the read-out electronic system for the Muon $g-2$ magnetic field scanner probes has a low-pass filter with a cut-off frequency at 90~kHz. The frequency-domain spectra of the noises in the FID and the phase function after smoothing are shown in Fig.~\ref{fig:TrolleyNoiseAmpSpectrum}. In these cases, the noise power spectrum is not a constant and thus the phase noise covariance matrix is not as simple as the form of Eq.~\ref{eq:phase_noise_cov}. If a large ensemble of noise waveforms are available, the corresponding phase noise can be calculated using Eq.~\ref{eq:phase_noise}, and the covariance matrix element $\Sigma_{jk}$ can be determined by calculating the ensemble average of $n_{\phi}(t_j)n_{\phi}(t_k)$. The ensemble of noise waveforms can be obtained by taking data with the magnetometer in a field outside its dynamic range and leaving the configurations of the electronics the same so that all sources of noise are included. Because the phase noise function $n_{\phi}(t)$ depends on the FID envelope and phase function, the noise covariance matrix has to be evaluated for each FID. Due to the filter effect, the high-frequency cut-off of $\tilde{n}_{\phi}(\omega)$ is much lower than $\pi/\Delta t-\omega_{0}$. Therefore, a larger down-sampling factor $\lambda$ is needed so that $\pi/(\lambda \Delta t)$ is smaller than the cut-off frequency of $\tilde{n}_{\phi}(\omega)$, and thus, the covariance matrix $\Sigma_{jk}$ becomes invertible. After obtaining an invertible noise covariance matrix, it can be used to construct the $\chi^{2}$ in the FID frequency extraction and statistical uncertainty determination. If the smoothing operation is performed, the shape of $\tilde{n}_{\phi}(\omega)$ for frequencies lower than the first zero position is similar to that for white noises (the red dashed curve in Fig.~\ref{fig:PhaseNoiseAmpSpectrum}) because the spectrum $\tilde{N}(\omega)$ is flat near $\omega_{0}$ (50~kHz). As more smoothing iterations are performed, the more similar these two spectra become. Many results of the studies performed for the white noise can be directly used for these measured FIDs with realistic noises.


In some cases, the noise spectrum may have sharp spikes at certain frequencies on top of a continuous spectrum. These peaks may be caused by electromagnetic interference with other devices. Suppose the single-frequency noise is $N(t)=N_{0}\cos(\omega_{N}t+\Phi_{0N})$ and the FID phase function is $\Phi(t)=\omega_{0}t$. According to Eq.~\ref{eq:phase_noise}, the phase noise is
\begin{align}
    n_{\phi}(t)=\frac{N_{0}}{A(t)}\sin((\omega_{N}-\omega_{0})t+\Phi_{0N}),
\end{align}
which is an oscillation at angular frequency $|\omega_{N}-\omega_{0}|$ that can be mitigated using the moving-average smoothing method. However, if $|\omega_{N}-\omega_{0}|$ is too small, the size of the smoothing window may be comparable to $T_{2}^{*}$ so the actual fit window will be very small after eliminating the edges. Therefore, noises with angular frequencies peaked near $\omega_{0}$ are almost irreducible. Furthermore, the polynomial fit of $\Phi(t)$ is affected more by low-frequency noise, particularly when $2\pi/|\omega_{N}-\omega_{0}|$ is longer than the fit window. For noises with sharp spikes in the frequency domain spectrum, the resolution depends on the FID frequency, and the resolution of the probe becomes significantly poorer when the FID frequency gets close to a noise frequency spike. 

\subsection{Unweighted and diagonal Minimum-$\chi^2$ Fit}
\label{subsec:unweighted_fit}

\begin{figure}[h!]
\centering
\includegraphics[width=0.7\linewidth]{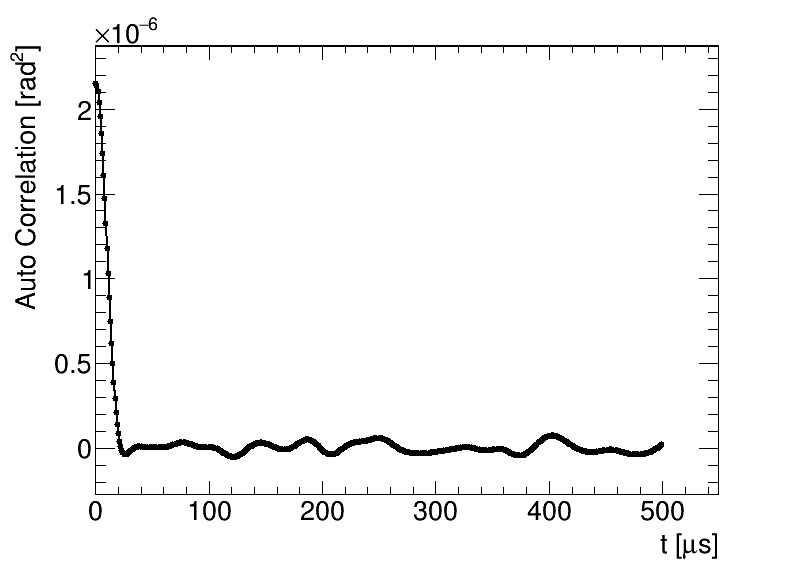}
\caption{Auto-correlation of samples in $n_{\phi}(t)$ for the noise of the Muon $g-2$ magnetic field scanner probe. \label{fig:AutoCorrelation}}
\end{figure}

Calculating the noise covariance matrix, particularly for the generic noise, is computation-intensive, so it is not suitable for online or large-scale FID analyses. Instead, the unweighted minimum-$\chi^2$ fit (assuming $\Sigma_{jk}\propto\delta_{jk}$) or the diagonal minimum-$\chi^2$ fit (keeping only diagonal elements of $\Sigma_{jk}$) are used if the minimized $\chi^{2}$-value is not used as a check of the goodness of the fit. The biases of the expectation and standard deviation of the extracted $\bar{\omega}$ are analysed using simulated FIDs with various envelope shapes, phase functions, and noise spectra. For both the unweighted and the diagonal fit, the fit result of $\bar{\omega}$ is always unbiased, and the standard deviations of $\bar{\omega}$ determined using these two methods are about 0 to 10\% larger than the fit result with the proper noise-correlation treatment described above. Therefore, if the unweighted or the diagonal fit is used, the fit result is not biased and the statistical uncertainty of the extracted $\bar{\omega}$ can be determined via the standard deviation of multiple measurements in the same field, but the $\chi^{2}/\nu$ cannot be used as an indicator of the goodness of the fit. However, if the down-sampling factor is significantly large, the diagonal fit generates the fit uncertainty and the minimal $\chi^{2}$ very close to those given by the fit with the correct noise covariance matrix. This effect can be explained using the auto-correlation spectrum of the smoothed $n_{\phi}(t)$ shown in Fig.~\ref{fig:AutoCorrelation}. For this $n_{\phi}(t)$, if two samples are separated by more than 20~$\mu$s, their auto-correlation is effectively zero. If the period after down-sampling is larger than 20~$\mu$s, the noise covariance matrix is essentially diagonal. This method can be applied when it is essential to obtain the statistical uncertainty from each FID and acceptable to worsen the statistical uncertainty with a sufficiently large down-sampling factor.

%% file: Conclusion.tex
\section{Conclusions}
\label{sec:conclusion}

We have presented a detailed prescription of implementing Cowan's method for extracting the FID frequency, which can be used in high-precision magnetometers. The phase function and the envelope function of an FID are determined using the Hilbert Transform. We have developed the methods for analyzing the effects caused by artifacts like the discrete Hilbert transform, baseline and signal distortion. These methods can be applied in the analysis frameworks of existing magnetometers to obtain systematic uncertainties, and they can also contribute to future designs of NMR read-out electronics to calculate the tolerances of artifacts. To leading order, these artifacts result in ripples in the phase function and the envelope function. A running-average method for smoothing the phase function has been developed to mitigate these effects. The remaining bias caused by the artifacts depends on their details, and it is recommended to use simulated FIDs with these artifacts to quantify their systematic biases on the frequency extraction.
In general, small $T_{2}^{*}$ times and large nonlinear terms in the phase function amplify these biases. 

Furthermore, the relationship between the noise in the phase function and the noise in the FID waveform has been derived to the leading order as shown in Eq.~\ref{eq:phase_noise}. The method for obtaining an invertible noise covariance matrix used in the minimal-$\chi^{2}$ fit has been described for the white and generic noise sources. The spectra of the noise in the phase functions have been discussed. The consistency between the statistical uncertainty generated by the minimum-$\chi^{2}$ fit and the standard deviation of the extracted FID frequency has been verified using simulated FIDs. This method is useful in determining the resolution of an NMR probe from a single shot when repeated measurements of the same field are difficult to achieve, and a $\chi^{2}$-test can be performed to determine the goodness of the fit. We have also verified that the fit results obtained through the unweighted and diagonal fits are not biased, and one can use these methods to extract the FID frequency without significantly worsening the resolution when the computing power is limited. 

%% file: Appendix.tex
\appendix

\section{Correlation between $n(t)$ and $n_{i}(t)$ samples for white noise}
\label{app:correlation_n_ni}

According to Eq.~\ref{eq:hilbert_transform_fourier}, the Hilbtert transform of an arbitrary noise function $N(t)$ is
\begin{align}
    N_{i}(t)&=\frac{1}{2\pi}\int_{0}^{\infty}\left[\int_{-\infty}^{\infty}N(\tau)(-i)\text{sgn}(\omega)e^{i\omega(t-\tau)}d\omega\right] d\tau \label{eq:Ni_Explicit}\\
    &=\frac{1}{2\pi}\int_{0}^{\infty}\left[\int_{-\infty}^{0}iN(\tau)e^{i\omega(t-\tau)}d\omega\right] d\tau
    +\frac{1}{2\pi}\int_{0}^{\infty}\left[\int_{0}^{\infty}(-i)N(\tau)e^{i\omega(t-\tau)}d\omega\right] d\tau \nonumber\\
    &=\frac{1}{2\pi}\int_{0}^{\infty}\left[\int_{0}^{\infty}iN(\tau)e^{-i\omega(t-\tau)}d\omega\right] d\tau
    +\frac{1}{2\pi}\int_{0}^{\infty}\left[\int_{0}^{\infty}(-i)N(\tau)e^{i\omega(t-\tau)}d\omega\right] d\tau \nonumber\\
    &=\frac{1}{\pi}\int_{0}^{\infty}\left[\int_{0}^{\infty}N(\tau)\frac{e^{i\omega(t-\tau)}-e^{-i\omega(t-\tau)}}{2i}d\omega\right] d\tau \nonumber\\
    &= \frac{1}{\pi}\int_{0}^{\infty}\left[\int_{0}^{\infty}N(\tau)\sin\left(\omega (t-\tau)\right)d\omega\right] d\tau . \nonumber
\end{align}
The integration over $\tau$ starts from 0 because the signal starts from $t=0$. Because of the finite sampling frequency, the integration over $\omega$ is truncated at $\pi/\Delta t$ where $\Delta t$ is the interval between samples. One can then simplify Eq~\ref{eq:Ni_Explicit} by performing the integration over $\omega$ and get
\begin{align}
    N_{i}(t)=P.V.\left\{\frac{1}{\pi}\int_{0}^{\infty}N(\tau)\frac{1-\cos(\frac{\pi}{\Delta t}(t-\tau))}{t-\tau} d\tau\right\}.\label{eq:Ni_derivation}
\end{align}
The principal value of the integral is taken, because the integrand of Eq.~\ref{eq:Ni_Explicit} is zero for $t=\tau$. Expressing the integral in Eq.~\ref{eq:Ni_derivation} as a sum over the discrete samples of $N(t)$, the $k$-th sample of $N_i(t)$ is 
\begin{align}
    N_{i}(t_{k})&=\frac{1}{\pi}\sum_{l=0, l\neq k}^{L-1} N(t_{l}) \frac{1-\cos(\frac{\pi}{\Delta t}(t_{k}-t_{l}))}{t_{k}-t_{l}} \Delta t \label{eq:Ni_expression} \\
    &=\frac{1}{\pi}\sum_{l=0, l\neq k}^{L-1} N(t_{l}) \frac{1-\cos(\pi(k-l))}{k-l} \nonumber\\
    &=\frac{1}{\pi}\sum_{l=0, l\neq k}^{L-1} N(t_{l}) \frac{1-(-1)^{k-l}}{k-l}, \nonumber
\end{align}
where $t_{k}=k\Delta t$ and $L$ is the total number of samples of $N(t)$.

If $N(t)$ is a white noise, any pair of different samples are statistically independent, and thus the expected value of $N(t_j)N(t_k)$ is
\begin{align}
    \left<N(t_j)N(t_k)\right>=\delta_{jk}\sigma^{2}_{N},
\end{align}
and then the expected value of $N(t_j)N_{i}(t_k)$ for $j\neq k$ is
\begin{align}
    \left<N(t_j)N_{i}(t_k)\right>&=\frac{1}{\pi}\sum_{l=0, l\neq k}^{L-1}  \left<N(t_{j})N(t_{l})\right> \frac{1-(-1)^{k-l}}{k-l}, \\
    &=\frac{1}{\pi}\sum_{l=0, l\neq k}^{L-1}  \delta_{jl}\sigma^{2}_{N} \frac{1-(-1)^{k-l}}{k-l}\nonumber\\
    &= \frac{1-(-1)^{k-j}}{\pi(k-j)} \sigma^{2}_{N}.\nonumber
\end{align}
For $j=k$, $\left<N(t_j)N_{i}(t_k)\right>=0$ because the $N_{i}(t_k)$ does not depend on $N(t_k)$ according to Eq.~\ref{eq:Ni_expression}. Finally, for $n(t)$ and $n_{i}(t)$ defined in Sec.~\ref{sec:noise}, the covariance matrix element between sample $j$ and $k$ is
\begin{align}
     \text{COV}(n(t_{j}),n_{i}(t_{k}))&=\left<n(t_{j})n_{i}(t_{k})\right>\\
     &=\frac{\left<N(t_{j})N_{i}(t_{k})\right>}{A(t_{j})A(t_{k})}\nonumber\\
     &=\frac{1-(-1)^{k-j}}{\pi(k-j)}\frac{\sigma_{N}^{2}}{A(t_{j})A(t_{k})}.\nonumber
\end{align}

\section{Correlation between different $n_i(t)$ samples for white noise}
\label{app:correlation_ni_ni}

The expected value of $N_{i}(t_j)N_{i}(t_k)$ can be calculated directly using Eq.~\ref{eq:Ni_expression}
\begin{align}
    \left<N_{i}(t_j)N_{i}(t_k)\right> &= \frac{1}{\pi^{2}}\left<\left[\sum_{m=0, m\neq j}^{L-1} N(t_{m}) \frac{1-(-1)^{j-m}}{j-m}\right] \left[\sum_{l=0, l\neq k}^{L-1} N(t_{l}) \frac{1-(-1)^{k-l}}{k-l}\right]\right> \label{eq:NiCorrelation}\\
    &=\frac{1}{\pi^{2}}\sum_{m=0, m\neq j}^{L-1}\sum_{l=0, l\neq k}^{L-1} \left< N(t_{m})N(t_{l})\right> \frac{1-(-1)^{j-m}}{j-m}\frac{1-(-1)^{k-l}}{k-l}\nonumber\\
    &=\frac{1}{\pi^{2}}\sum_{m=0, m\neq j}^{L-1}\sum_{l=0, l\neq k}^{L-1} \delta_{ml}\sigma^{2}_N \frac{1-(-1)^{j-m}}{j-m}\frac{1-(-1)^{k-l}}{k-l}\nonumber\\
    &=\frac{\sigma^{2}_N}{\pi^{2}}\sum_{m=0, m\neq j,m\neq k}^{L-1} \frac{(1-(-1)^{j-m})(1-(-1)^{k-m})}{(j-m)(k-m)}.\nonumber
\end{align}

If $j$ and $k$ are not close to the ends (0 or L), then the bounds of the sum in Eq.~\ref{eq:NiCorrelation} can be extended to $(-\infty,+\infty)$. The following discussions assume that $j$ and $k$ are not close to the ends. Particularly, for $j=k$, Eq.~\ref{eq:NiCorrelation} becomes
\begin{align}
    \left<N_{i}^{2}(t_j)\right> &= \frac{\sigma^{2}_N}{\pi^{2}}\sum_{m=-\infty, m\neq j}^{\infty} \frac{(1-(-1)^{j-m})^{2}}{(j-m)^{2}}\label{eq:NiSelfCorrelation}\\
    &=2\frac{\sigma^{2}_N}{\pi^{2}}\sum_{m=1}^{\infty}\frac{4}{(2m+1)^{2}}\nonumber\\
    &=8\frac{\sigma^{2}_N}{\pi^{2}}\frac{\pi^{2}}{8}\nonumber\\
    &=\sigma^{2}_N.\nonumber
\end{align}
According to Eq.~\ref{eq:NiCorrelation}, it is obvious that $\left<N_{i}(t_j)N_{i}(t_k)\right>=0$ when $j-k$ is an odd number. When $j-k$ is an even number, Eq.~\ref{eq:NiCorrelation} can be further simplified as
\begin{align}
    \left<N_{i}(t_j)N_{i}(t_k)\right>  &=\frac{\sigma^{2}_N}{\pi^{2}}\sum_{m=-\infty, m\neq j,m\neq k}^{\infty} \frac{(1-(-1)^{j-m})^{2}}{(j-m)(k-m)}\\
    &=\frac{\sigma^{2}_N}{\pi^{2}}\sum_{m=-\infty, m\neq j,m\neq k}^{\infty} \frac{(1-(-1)^{j-m})^{2}}{k-j}\left[\frac{1}{j-m}-\frac{1}{k-m}\right]\nonumber\\
    &=0\nonumber
\end{align}
Therefore, the covariance matrix element between sample $j$ and $k$ of $n_{i}(t)$ is 
\begin{align}
     \text{COV}(n_{i}(t_{j}),n_{i}(t_{k}))&=\left<n_{i}(t_{j})n_{i}(t_{k})\right>\\
     &=\frac{\left<N_{i}(t_{j})N_{i}(t_{k})\right>}{A(t_{j})A(t_{k})}\nonumber\\
     &\approx\frac{\sigma^{2}_N}{A^{2}(t_{j})}\delta_{jk},\nonumber
\end{align}
where the approximation depends on how far $j$ and $k$ are from the ends. One can use Eq.~\ref{eq:NiCorrelation} to compute $\left<N_{i}(t_{j})N_{i}(t_{k})\right>$ and $\left<n_{i}(t_{j})n_{i}(t_{k})\right>$ accurately. For example, $\left<N_{i}(t_{j})N_{i}(t_{k})\right>=0.5\sigma^{2}_N$ for $j=k=0$, while $\left<N_{i}(t_{j})N_{i}(t_{k})\right>=0.975\sigma^{2}_N$ for $j=k=7$. 

\section{Rank of the covariance matrix}
\label{app:rank_covariance_matrix}

Let $L$ be the total number of samples in the fit region. The dimension of the $\Sigma$ matrix in Eq.~\ref{eq:phase_noise_cov} is $L\times L$. $\Sigma$ is invertible if and only if its rank is L, or equivalently, linear equations
\begin{align}
    \Sigma X = 0
    \label{eq:sigma_equations}
\end{align}
have only one solution $X=0$. By definition, 
\begin{align}
    \Sigma = \left< n_{\phi}n_{\phi}^{T} \right>,
\end{align}
where $n_{\phi}$ is a column-vector representing the phase noise in the fit region. The Fourier transform of $n_{\phi}(t)$ expressed as a vector product is
\begin{align}
    \tilde{n}(\omega)=n_{\phi}^{T}Z(\omega),
\end{align}
where $Z(\omega)$ is a column-vector with the $j$-th element  $Z_{j}(\omega)=\exp(-i\omega(j\Delta t))$. According to Eq.~\ref{eq:spectrum_constraint}, for $\omega>\pi/\Delta t-\omega_{0}$ approximately
\begin{align}
    \tilde{n}(\omega)=n_{\phi}^{T}Z(\omega)=0.
    \label{eq:constraints_derivation}
\end{align}
After multiplying $n_{\phi}$ to both sides of Eq.~\ref{eq:constraints_derivation} and taking the expected value, one gets
\begin{align}
    \left< n_{\phi}n_{\phi}^{T} \right>Z(\omega)=\Sigma Z(\omega)=0.
    \label{eq:constraints_derivation2}
\end{align}
Because of the discrete Fourier transform, $\omega$ can only be an integer times $\Delta \omega=2\pi/T$ up to the Nyquist angular frequency $\pi/\Delta t$, where $T$ is the duration of the entire signal. In the region $(\pi/\Delta t-\omega_{0},\pi/\Delta t)$, there are $\omega_{0}/\Delta \omega=\omega_{0} T/2\pi$ values of $\omega$ that satisfies Eq.~\ref{eq:constraints_derivation2}. In other words, there are $\omega_{0} T/2\pi$ non-trivial solutions to the linear equations in Eq.~\ref{eq:sigma_equations}. The rank of matrix $\Sigma$ is then $L-\omega_{0} T/2\pi$, and thus, $\Sigma$ is not intervible.